\documentclass[10pt,conference,draft]{IEEEtran}
\usepackage{amssymb,amsmath,palatino}

\begin{document}
\title{On the Decoding Complexity of \\ Cyclic Codes Up to the BCH Bound}
\author{
\authorblockN{Davide Schipani}
\authorblockA{Mathematics Institute\\
University of Z\"urich\\
CH-8057 Z\"urich\\
davide.schipani@math.uzh.ch}
\and
\authorblockN{Michele Elia}
\authorblockA{Dipartimento di Elettronica\\
 Politecnico di Torino\\
  IT-10129, Torino\\
elia@polito.it}
\and
\authorblockN{Joachim Rosenthal}
\authorblockA{Mathematics Institute\\
University of Z\"urich\\
CH-8057 Z\"urich\\
http://www.math.uzh.ch/aa}

}
\maketitle

\begin{abstract}
\noindent
The standard algebraic decoding algorithm of cyclic codes $[n,k,d]$ up to the
 BCH bound $\delta=2t+1$ is very efficient and practical for relatively small $n$ 
 while it becomes unpractical for large $n$ as its computational complexity
 is $O(nt)$. 
Aim of this paper is to show how to make this algebraic decoding computationally more efficient:
 in the case of binary codes, for example, the complexity of the syndrome computation drops from $O(nt)$ to $O(t\sqrt n)$,
 while the average complexity of the error location drops from $O(nt)$ to 
$\max \{ O(t\sqrt n), O(t\log^2(t)\log\log(t)\log(n)) \}$. 
\end{abstract}

\vspace{2mm}

\vspace{2mm}



\section{Introduction}
The algebraic decoding of cyclic codes  up to the BCH bound, as      
 obtained early in the sixties with the contribution of many people,
 was considered very efficient for the needs of that time (\cite{berlekamp,blahut,sloane,massey,peterson,pless}). 
However, today we can and need to manage error correcting codes of sizes that
 require more efficient algorithms, possibly at the limit of their theoretical
 minimum complexity. We are proposing here an algorithm that goes in this direction. 

\noindent Although we will focus as our main point of reference and comparison on the classical algebraic decoding, there are other decoding algorithms that have been recently proposed and that we limit ourselves to cite here as a reference, e.g. \cite{fitz, guerrini, mora, orsini}.

\noindent
Let us summarize now 
the standard algebraic decoding of cyclic codes:
 let $\mathcal C$ be an $[n,k,d]$ cyclic code over a finite field $\mathbb F_q$, $q=p^s$ for a prime $p$, with generator
 polynomial of minimal degree $r=n-k$
$$ g(x)= x^r+g_1 x^{r-1}+ \ldots +g_{r-1}x +g_r ~~, $$ 
$g(x)$ dividing $x^n-1$, and let $\alpha$ be a primitive $n$-th root of unity lying in a finite field $\mathbb F_{p^{m}}$, where the
extension degree is the minimum integer $m$ such that $n$ is a divisor of $p^m-1$.
Assuming that $\mathcal C$ has BCH bound $\delta=2t+1$ (if $\delta$ is even, we would just consider $\delta-1$), then $g(x)$ has $2t$ roots with
 consecutive power exponents, so that the whole set of roots is 
$$\mathfrak R=\{ \alpha^{\ell+1}, \alpha^{\ell+2}, \ldots, \alpha^{\ell+2t}, 
                 \alpha^{s_{2t+1}}, \ldots , \alpha^{s_{r}}   \} ~~, $$
where it is not restrictive to take $\ell=0$ as it is usually done. \\
%
Let $R(x)=g(x)I(x)+e(x)$  be a received code word such that the error pattern $e(x)$
 has no more than $t$ nonzero coefficients. The Gorenstein-Peterson-Zierler decoding procedure
 (\cite{sloane,peterson}), which is a standard decoding procedure for every cyclic code
 up to the BCH bound, is made up of four steps:
\begin{itemize}
  \item Computation of $2t$ syndromes: $S_j=R(\alpha^{j})$, $j=1,\ldots ,2t$.
  \item Computation of the error-locator polynomial   
   $\sigma(z)= \sigma_t z^t+\sigma_{t-1} z^{t-1}+ \cdots +\sigma_{1}z+1$ (we are assuming the case that
    exactly $t$ errors occurred; if there are $t_e<t$ errors, this step would output a polynomial
    of degree $t_e$). 
  \item Computation of the roots of $\sigma(z)$ in the form $\alpha^{-j_h}$, $h=1, \ldots, t$,
   yielding the error positions $j_h$.
  \item Computation of the error magnitudes. 
\end{itemize}

\noindent
Efficient implementations of this decoding algorithm combine the computation of
 $2t$ syndromes using  Horner's rule, the Berlekamp-Massey algorithm to obtain
 the error-locator polynomial, the Chien search to locate the errors,  and
 the evaluation of Forney's polynomial to estimate the error magnitudes. 

\noindent
The computation of the $2t$ syndromes using Horner's rule requires $2tn$ multiplications in $\mathbb F_{p^m}$, which may be prohibitive when $n$ is large.
The Berlekamp-Massey algorithm has multiplicative complexity
 $O(t^2)$ (\cite{blahut,vetterli}), is very efficient and will not be discussed further later on. The Chien search requires again $O(tn)$
 multiplications in $\mathbb F_{p^m}$ and Forney's algorithm $O(t^2)$ (\cite{vetterli}). Notice that this fourth step is not required if we deal with binary codes and that both the first and the fourth steps consist primarily in polynomial evaluations, so they can benefit from any efficient polynomial evaluation algorithm, as we will show.

\noindent
The standard decoding procedure is satisfactory when the code length $n$ is not too large
 (say $< 10^3$) and efficient implementations are set up taking advantage of the
 particular structure of the code.
The situation changes dramatically when $n$ is of the order of $10^6$ or larger.
 In this case a complexity $O(tn)$, 
 required by the syndrome evaluations and by the Chien search, is not acceptable anymore. 

\noindent This paper describes some methods to make these steps more efficient and practical even for large $n$. We will follow the usual approach of focusing as above in computing the number of multiplications, as they are more expensive than sums (see also \cite{elia}). 

\noindent 
The paper is structured as follows: Section II 
 concerns the computation of syndromes. 
Section III deals with the computation of the roots of the error-locator polynomial as well as the corresponding error
 positions; the error locator polynomial is supposed to be given (being computed by Berlekamp-Massey algorithm). 
Finally, Section IV 
gives a numerical example illustrating the whole procedure. 


 \section{Syndrome Evaluation}

 Let $\beta$ be any element of $\mathfrak R$, the standard Horner's rule (\cite{Jungnickel},\cite{knuth2}) allows us to compute 
 $R(\beta)$ in at most $n$ products, thus for the computation of $2t$ syndromes we have
 the estimate $O(tn)$. 
However, in \cite{schip5,schip4} we showed that polynomials over a finite field of characteristic $p$ 
 can be evaluated more efficiently by exploiting the Frobenius automorphism, i.e. 
 the mapping $\sigma(\beta) = \beta^p$, with a significant computational cost reduction. 

\noindent
Briefly, to evaluate a polynomial $r(x)$ of degree $n$ 
over $\mathbb F_{p^s}$, in $\beta$, an element of $\mathbb F_{p^m}$, we write $r(x)$ 
 as a linear combination of $s$ polynomials $r_i(x)$ over $\mathbb F_{p}$
$$
r(x) = r_0(x)+ \gamma r_1(x)+\cdots +\gamma^{s-1} r_{s-1}(x) ~~,
$$
 where $\{1, \gamma, \ldots , \gamma^{s-1} \}$ is a basis for $\mathbb F_{p^s}$. 
 Thus $r(\beta)$ is obtained as a linear combination of $s$ field elements $r_i(\beta)$.
To evaluate a polynomial $R(x)$ over $\mathbb F_{p}$ in $\beta$, one can profit by writing 
$$
R(x)= R_{1,0}(x^p)+ x R_{1,1}(x^p)+\cdots +x^{p-1} R_{1,p-1}(x^p) ~~,
$$
where $R_{1,0}(x^p)$ collects the powers of $x$ with exponent a multiple
of $p$ and in general $x^{i} R_{1,i}(x^p)$ collects the powers of the form
$x^{ap+i}$, with $a\in\mathbb{N}$ and $0\leq i\leq p-1$. Thus $R(\beta)$ can be computed
 by evaluating $\beta^p$, then computing every $R_{1,i}(\beta^p)$ and finally computing the linear
 combination.
This procedure requires, for example, nearly $n/2$ multiplications in the binary case, but the
 further advantage is that it can be iterated. After $L$ steps, we need to evaluate $p^L$ polynomials
 $R_{L,i}(x)$ of degree at most $\lfloor \frac{n}{p^L} \rfloor$.
 By a convenient number $L$ of iterations, and with a smart arrangement
 of the multiplications (\cite{schip5}), one can achieve an overall complexity of approximately
 $2s\sqrt{n(p-1)}$. In the particular case of binary codes, the complexity is
 $2\sqrt{n}$. 

\vspace{5mm}
\noindent
It should be remarked that in hardware implementations, the proposed algorithm allows a strong parallelism,
 while Horner's rule is inherently serial. In fact, if $L$ is the number of iterations, the evaluation
 of the $p^L$ polynomials $R_{L,i}(x)$ can be done in parallel. 
Moreover an additional gain may be given by the
 pre-computation of the powers of $\beta$, especially when the number of syndromes to be computed is big. Furthermore, like in Horner's rule,
 multiplication by $\beta$ or its powers can be performed using Linear Feedback Shift Registers (\cite{golomb,sloane,mcel1})
 with a further speed up at a very small cost, while the $p$-power operations would benefit from
 the use of a normal basis (\cite{Jungnickel,lidl}).

\vspace{5mm}
\noindent 
Lastly, it should be also remarked that, in particular situations, a better cost reduction
 can be obtained by means of a different use of the Frobenius automorphism and a careful
 choice of the number of iterations.  As an example, in \cite{schip4} we described 
 a method of computing the syndromes for the famous Reed-Solomon code $[255,223,33]$ over
 $\mathbb F_{2^{8}}$ used by NASA (\cite{wicker2}), that employs $6735$ multiplications to evaluate $32$ syndromes, versus $8159$ multiplications that are necessary using Horner's rule.
 The direct application of the method outlined above would not  be convenient in this situation because of the particular parameters involved.

\section{Roots of the error-locator polynomial}

Once the error locator polynomial $\sigma(z)$ is computed from the syndromes
using the Berlekamp-Massey algorithm, its roots, represented in the form
 $\alpha^{-\ell_i}$, correspond to the error positions $\ell_i$, $i=1, \ldots, t$,
 which are generally found by testing $\sigma(\alpha^{-i})$ for all $n$ possible powers 
 $\alpha^{-i}$ with an algorithm usually referred to as the Chien search.
In this approach, if $\sigma(\alpha^{-j})=0$ an error in position $j$ is recognized,
 otherwise the position is correct. However, this simple mechanism can be
 unacceptably slow when $n$ is large 
 since its complexity is $O(t n)$: aim of this Section is to describe a less costly procedure.\\
The Cantor-Zassenhaus probabilistic factorization algorithm (\cite{cantor}) is
 very efficient in factoring a polynomial and consequently in computing the
 roots of a polynomial (\cite{benor,jurgen}).
Since $\sigma(z)$ is the product of $t$ linear factors $z+\rho_i$
  over $\mathbb F_{p^m}$ (i.e. $\rho_i$ is a $p$-ary polynomial in $\alpha$
  of degree $m-1$), this factoring algorithm can be directly applied to separate
  these $t$ factors. The error positions
  $\ell_i$  are then obtained by computing the discrete logarithm of $(\rho_i)^{-1}=\alpha^{\ell_i}$
  to base $\alpha$. This task can be performed by Shank's algorithm (\cite{sh71}), which we revisit
  below. The overall expected complexity of finding the error positions with this algorithm
  is  $O(mt \log^2 t\log\log t)$ (\cite{benor}), plus $O(t\sqrt n)$, where the second
  addend comes from Shank's algorithm. 
Moreover, better computational estimates may be obtained taking into account the considerations and improvements highlighted in \cite{schip2}. 


\paragraph{Cantor-Zassenhaus algorithm}
The  algorithm of Cantor-Zassenhaus (\cite{cantor}) is described here
 for easy reference (see also the analysis in \cite{schip2}). We describe only the case of characteristic $2$, which is by far the most common in practice; the interested reader can find the general situation in \cite{cantor,schip2}.
Assume that $p(z)$ is a polynomial over $\mathbb F_{2^m}$ that is a product of $t$
  polynomials of degree $1$ over the same field $\mathbb F_{2^m}$, $m$ even (when $m$ is odd it is enough to
  consider a quadratic extension and proceed as in the case of even $m$). Suppose that $\alpha$ is
  a known primitive element in $\mathbb F_{2^m}$, and set $\ell_m=\frac{2^m-1}{3}$, then 
  $\rho=\alpha^{\ell_m}$ is a primitive cubic root in $\mathbb F_{2^m}$, so that $\rho$
  is a root of $z^2+z+1$.
The algorithm consists of the following steps:

 \begin{enumerate}
    \item Generate a random polynomial $b(z)$ of degree not greater than $t-1$ over $\mathbb F_{2^m}$. 
    \item Compute $a(z)=b(z)^{\ell_m} \bmod p(z)$.
    \item IF $a(z) \neq 0,1,\rho,\rho^2$,  THEN at least a polynomial among 
\begin{multline*}
     \gcd\{p(z), a(z)\}, \gcd\{p(z), a(z)+1 \},\\ \gcd\{p(z), a(z)+\rho \}, \gcd\{p(z), a(z)+\rho^2 \}
 \end{multline*}    
     will be a non trivial factor of $p(z)$,  ELSE repeat from point 1.
    \item Iterate until all linear factors of $p(z)$ are found.
 \end{enumerate}

\paragraph{Remark 1} As shown in \cite{schip2}, the polynomial $b(z)$ can be conveniently chosen of the form $z+\beta$,
  using $b(z)=z$ as initial choice. Let $\theta$ be a generator of the cyclic subgroup of $\mathbb F_{2^m}^*$ of order 
  $\ell_m$.  
If $z^{\ell_m}=\rho^i \bmod \sigma(z)$, $i \in \{0,1,2 \}$, then each root $\zeta_h$
 of $\sigma(z)$ is of the form $\alpha^i \theta^j$. 
If this is the case, which does not allow us to find a factor, we repeat the test with $b(z)=z+\beta$
 for some $\beta$ and we will succeed as soon as the elements $\zeta_h+\beta$ are not all of the type
 $\alpha^i \theta^j$ for the same $i \in \{0,1,2\}$.
This can be shown to happen probabilistically, and often deterministically, very soon, expecially when the degree of $\sigma(z)$ is high. In most practical situations it is actually very seldom that more than two iterations are needed, which explains its widespread use.

\paragraph{Shank's algorithm} 
Shank's algorithm can be applied to compute the discrete logarithm in a group
  of order $n$ generated by the  primitive element $\alpha$. The exponent $\ell$ in the 
  equality
$$    \alpha^{\ell} = b_0 + b_1 \alpha + \cdots + b_{s-1} \alpha^{s-1} ~~. $$
 is written in the form $\ell = \ell_0 + \ell_1 \lceil \sqrt n \rceil$.
A table $\mathcal T$ is constructed with $\lceil \sqrt n \rceil$ entries
 $\alpha^{\ell_1 \lceil \sqrt n \rceil}$ which are sorted in some well defined order,
 then a cycle of length $\lceil \sqrt n \rceil$ is started computing 
$$ A_j= ( b_0 + b_1 \alpha + \cdots + b_{s-1} \alpha^{s-1}) \alpha^{-j} 
   ~~j=0, \ldots , \lceil \sqrt n \rceil-1 ~~, $$
  and looking for $A_j$ in the Table; when a match is found with the $\kappa$-th entry,
  we set $\ell_0=j$ and $\ell_1=\kappa$, and the discrete logarithm $\ell$
  is obtained as $j+\kappa \lceil \sqrt n \rceil$. \\
This algorithm can be performed with complexity $O(\sqrt{n})$ both in time and 
  space (memory). In our scenario, since we need to compute $t$ roots, the complexity
  is $O(t\sqrt n)$.

\paragraph{Remark 2} 
The Cantor-Zassenhaus algorithm finds the roots $X_j=\alpha^{\ell_j}$ of the
 reciprocal of the error locator polynomial, then the baby-step giant-step algorithm of Shank's 
finds the error positions $\ell_j$s. 
As said in the introduction, this is the end of the decoding process for binary codes.
For non-binary codes, Forney's polynomial $\Gamma(x)=\sigma(x) (S(x)+1) \bmod x^{2t+1}$,
 where $S(x)=\sum_{i=1}^{2t} S_i x^i$  (\cite{wicker}), yields the error values 
$$  Y_j =-X_j \frac{\Gamma(X_j^{-1})}{\sigma'(X_j^{-1})} ~~.   $$
Again we remark that this last step can benefit from an efficient polynomial evaluation algorithm, such as the one discussed in Section 2.

\paragraph{Remark 3} 
We observe that the above procedure can be used to decode beyond
 the BCH bound, up to the minimum distance, whenever the error locator polynomial
 can be computed from a full set of syndromes (\cite{elia1,truong,reed,wicker}).



\section{A numerical example}
In the previous sections we presented methods to compute syndromes and error locations
 in the GPZ decoding scheme of cyclic codes up to their BCH bound, which are asymptotically
 better than the classical algorithms.  
%
%
The following example illustrates the complete new procedure. 
 
%

\noindent Consider a binary BCH code $[63,45,7]$ with generator polynomial
  $$g(x)= x^{18}+x^{17}+x^{14}+x^{13}+x^9+x^7+x^5+x^3+1 $$
whose roots are 
\begin{multline*}
\alpha, \alpha^{2},  \alpha^{4}, \alpha^{8}, \alpha^{16}, \alpha^{32},
 \alpha^{3}, \alpha^{6}, \alpha^{12}, \\ 
\alpha^{24}, \alpha^{48}, \alpha^{33}, \alpha^{5},
  \alpha^{10}, \alpha^{20}, \alpha^{40}, \alpha^{17}, \alpha^{34},
\end{multline*}
  thus the BCH bound is $7$. 
Let $c(x)=g(x) I(x)$ be a transmitted code word, and the received word be
\begin{multline*}
 r(x) = x^{57}+x^{56}+x^{53}+x^{52}+x^{50}+x^{48}+x^{46}+x^{44}+x^{42}+\\x^{39}+ 
    x^{31}+x^{18}+x^{17}+x^{14}+x^{13}+x^{7}+x^5+x^3+1 
\end{multline*}
where $3$ errors occurred. The $6$ syndromes are
$$   \left\{  \begin{array}{l} 
    S_1 = \alpha^5+\alpha^2+\alpha \\ 
    S_2 = S_1^2  \\ 
    S_3= \alpha^5+\alpha^4+\alpha^3+\alpha^2+\alpha \\ 
    S_4= S_1^4 \\ 
    S_5= \alpha^5+\alpha^2+1 \\ 
    S_6= S_3^2 \\ 
     \end{array} \right. ~~.
$$ 
For example, $S_1$ has been computed considering $r(x)$ as 
$$
[r_{3,0}+zr_{3,1}+y(r_{3,2}+zr_{3,3})]+x[r_{3,4}+zr_{3,5}+y(r_{3,6}+zr_{3,7})],
$$
with $y=x^2$, $z=x^4$, $w=x^8$ and
$$   \left\{  \begin{array}{l} 
    r_{3,0} = w^{7}+w^{6}+1 \\     
     r_{3,1} = w^{6}+w^{5} \\     
    r_{3,2} = w^{6}+w^{5}+w^{2}  \\ 
     r_{3,3} = w^{5}+w  \\ 
    r_{3,4} = w^{7}+w^{2}  \\ 
     r_{3,5} = w^{6}+w+1  \\ 
    r_{3,6} = 1  \\ 
     r_{3,7} = w^{4}+w^{3}+1  \\ 
     \end{array} \right. ~~
$$ 

with only $16$ products, namely $3$ to compute $\alpha^2$, $\alpha^4$ and $\alpha^8$, $6$ for the powers of $w$ up to $w^7$ and $7$ multiplications by $x$, $y$ and $z$.

The coefficients of the error locator polynomial turn out to be
$$  \left\{  \begin{array}{l} 
      \sigma_1 = \alpha^5+\alpha^2+\alpha \\ 
      \sigma_2 = \alpha^3+\alpha^4+\alpha \\ 
      \sigma_3 = \alpha^4+\alpha^5+\alpha^2 \\ 
   \end{array} \right. ~~.
$$ 
The roots of $\sigma^*(z)=z^{3}\sigma(z^{-1})=\prod_{i=1}^3(z-\alpha^{\ell_i})$ are computed  as follows using the
  Cantor-Zassenhaus algorithm. 
  
Let $\rho=\alpha^{21}$ be a cube root of the unity; consider a random polynomial, for
  instance $z+\rho$, of degree less than $3$ and compute $a(z)=(z+\rho)^{21}$ modulo
  $\sigma^*(z)$ (the exponent of $z+\rho$ is $\frac{2^m-1}{3}=\frac{63}{3}=21$):
$$ (\alpha^5+\alpha^4+\alpha^2+\alpha+1) z^2+(\alpha^3+\alpha+1) z+\alpha^5+\alpha^4+x^3+1 ~~.  $$
In this case $a(z)$ has no root in common with $\sigma^*(z)$, while 
 \begin{multline*}
\gcd(a(z)+1,\sigma^*(z))=z+(\alpha^4+\alpha^3+1)\ \ \ \ (\ell_1=31),  \\
\gcd(a(z)+\rho,\sigma^*(z))=z+( \alpha^5+\alpha^4+\alpha^2+1)\ \ \ (\ell_2=9), \\ 
\gcd(a(z)+\rho^2,\sigma^*(z))=z+( \alpha^3+\alpha)\ \ \ \ (\ell_3=50).
 \end{multline*}

The error positions have been obtained using Shank's algorithm with a table of $8$
 entries, and a loop of length $8$ for each root, for a total of $24$ searches versus
 $63$ searches of Chien's search. 


 


\section{Concluding remarks}

A new decoding algorithm for cyclic codes has been presented having a very competitive complexity and targeting in particular those applications using error correcting codes with very large length. 

\section*{Acknowledgment}
We thank the anonymous referees for valuable comments that have allowed to improve and clarify the formulation of the paper.
The Research was supported in part by the Swiss National Science
Foundation under grant No. 132256. 

\IEEEtriggeratref{26}


\end{document}